\documentclass{PoS}

\usepackage[intlimits]{amsmath}
\usepackage{amssymb}
\usepackage{mathrsfs}
\usepackage{dsfont}
\usepackage{subfigure}

\title{The QCD Phase Diagram with Effective Theories}

\ShortTitle{QCD Phase Diagram with Effective Theories}

\author{\speaker{Ydalia Delgado}
\\Institut f\"ur Physik, Karl-Franzens Universit\"at, Graz, Austria
\\E-mail: \email{ydalia.delgado-mercado@uni-graz.at}}

\author{Hans Gerd Evertz
\\Institute for Theoretical and Computational Physics, 
Technische Universit\"at Graz, Austria
\\E-mail: \email{evertz@tugraz.at}}

\author{Christof Gattringer
\\Institut f\"ur Physik, Karl-Franzens Universit\"at, Graz, Austria
\\E-mail: \email{christof.gattringer@uni-graz.at}}

\author{Daniel G\"oschl
\\Institut f\"ur Physik, Karl-Franzens Universit\"at, Graz, Austria
\\E-mail: \email{daniel.goeschl@aon.at}}

\abstract{
We study two effective theories for QCD at non-zero temperature and finite chemical
potential, using local Polyakov loops as the degrees of freedom. The sign problem  is
solved by exactly mapping the partition function to a sum over flux and monomer
variables with only real and positive weights, making the two models accessible to
Monte Carlo simulation techniques. We use generalized worm algorithms and a local
Metropolis update to perform  the simulations and determine the phase diagram as a
function of the temperature and the chemical potential.
}

\FullConference{XXIX International Symposium on Lattice Field Theory \\
                 July 10 $-$ 16 2011\\
                 Squaw Valley, Lake Tahoe, California}
 
\begin{document}

\section{Introductory remarks}
\vspace{-1mm}
\noindent Running and upcoming experiments, e.g., ALICE, RHIC, FAIR and NICA will give
access to different states of strongly interacting matter at various temperatures and
densities. Mapping these states in the QCD phase diagram is a challenging task that
requires understanding of non-perturbative QCD since non-perturbative processes
dominate in the transition regions. In principle lattice QCD is a powerful tool  to
address non-perturbative phenomena quantitatively, as has, e.g., been demonstrated in
studies of the phase diagram at zero density \cite{zeromu}. At finite density,
however,  the fermion determinant becomes complex making QCD inaccessible to Monte
Carlo simulations already at moderate densities. Only regions of small chemical potential
can be studied on the lattice using methods which extrapolate from zero density, such
as reweighting techniques, Taylor expansion, imaginary chemical potential, et
cetera (see
\cite{other} for recent reviews), while the rest of the phase diagram remains
unexplored and requires new ideas.

\vspace*{2mm} \noindent In this work we explore effective theories which can be
derived from full QCD using the strong coupling approximation for the gluon action
and a hopping expansion for the fermion determinant. Concerning the gluon interaction
the two models are based on the 
relation of center symmetry and the deconfiment transition \cite{znbreaking}. In
addition they take into account the leading center symmetry breaking terms which couple
to the chemical potential $\mu$. At $\mu \ne 0$ these terms give rise to a complex
action and the models inherit the complex phase problem of QCD. We explore two
variants of the model: The {\sl ``$SU(3)$ effective theory''} \cite{karschwyld}, 
where the remaining degrees of freedom are traced SU(3) valued spins, and a second
version where the spins are further reduced to the center group $\mathds{Z}_3 = 
\{ 1, e^{i2\pi/3}, e^{-i2\pi/3}\}$. The latter we refer to as the 
{\sl ``$\,\mathds{Z}_3$ effective theory''} \cite{flux}. In both cases one can
solve the complex phase problem by an exact mapping onto a flux representation 
(see \cite{cg} for the SU(3) case, and \cite{flux} for the $\mathds{Z}_3$ model), where
only real and non-negative terms appear in the partition sum. In this form Monte Carlo
simulations are possible for arbitrary chemical potential and we use generalized 
Prokof'ev-Svistunov worm algorithms \cite{worm} and local Metropolis updates to study
the phase diagram. Our results shed light on the role of center symmetry in the QCD
phase diagram \cite{ours}. For the SU(3) case the model is accessible to 
complex Langevin techniques \cite{karschwyld,langevin} and our results from the flux
representation may also serve as reference data to study that approach.

\section{Effective theories}
\vspace{-1mm}
\noindent In pure gauge theory the confinement-deconfinement transition is
related to the spontaneous breaking of center symmetry \cite{znbreaking} and the
Polyakov loop, which represents a static quark source, can be used as an order
parameter. When matter fields are coupled, center symmetry is broken explicitly
by the fermion determinant. However, one may expect that the underlying symmetry
still governs parts of the dynamics of the full theory.  Thus we consider
effective theories with both center symmetric and center symmetry breaking
terms.  The action has the general form

\begin{equation}
S \;  = \; - \!\sum_x \left(\! \tau \! \sum_{\nu = 1}^3 \! \Big[ L(x) L(x+\hat{\nu})^\star
+ c.c. \Big] 
+ \kappa \Big[ e^\mu L(x) +  e^{-\mu} L(x)^\star \Big]\! \right)\; .
\label{action_su3} 
\end{equation}

\noindent In the SU(3) effective theory the degrees of freedom $L(x)$ are the traced
SU(3) variables $L(x) =$ Tr $P(x)$ with $P(x) \in$ SU(3) attached to the sites
$x$  of a three-dimensional cubic lattice which we consider to be finite with
periodic boundary conditions. By $\hat{\nu}$ we denote the unit vector in
$\nu$-direction, with $\nu = 1, 2, 3$.  The first term of the action, i.e., the
nearest neighbor term, can be obtained as the leading contribution in the strong
coupling expansion of the effective action for the Polyakov loop. This term is invariant under center
transformations $L(x) \rightarrow z L(x)$ with $z \in \mathds{Z}_3$.  The parameter
$\tau$ depends on the temperature (it increases with $T$) and is real and positive.
The second term, referred to as the magnetic term, is obtained as the leading
$\mu$-dependent contribution in the hopping expansion (large mass expansion) of the
fermion determinant.  The real and positive  parameter $\kappa$ is proportional to
the number of flavors and  depends on the fermion mass (it decreases with $m_q$).
The magnetic  term breaks center symmetry explicitly and is complex when the
chemical potential $\mu$ is non-zero, thus generating a complex phase problem.

\vspace*{2mm}
\noindent The grand canonical partition function of the model described by
(\ref{action_su3}) is obtained by integrating the Boltzmann factor $e^{-S[P]}$
over all configurations of the Polyakov loop variables. The corresponding measure is
a product over the reduced Haar measures $dP(x)$ at the sites $x$. Thus

\begin{equation}
Z \; = \prod_x \int_{SU(3)} dP(x) \, e^{-S[P]} \; = \; \int D[P] \, e^{-S[P]} \; .
\label{sum_su3} 
\end{equation}
Equations (\ref{action_su3}) and (\ref{sum_su3}) define the SU(3) effective
theory. Exploring the Yaffe-Svetitsky conjecture \cite{znbreaking}, it is possible to simplify the
effective theory further by using spin variables $p_x \in \mathds{Z}_3$.  
The action of the resulting $\mathds{Z}_3$ effective theory is given
by

\begin{equation}
S[p]  \; = \; - \!\sum_x \left(\! \tau \! \sum_{\nu = 1}^3 \! \Big[ p_x p_{x+\hat{\nu}}^* + c.c. \Big] 
+ \kappa \Big[ e^\mu p_x +  e^{-\mu} p_x^* \Big]\! \right)\ ,
\label{action_z3} 
\end{equation}
and the partition function is a sum over all possible spin configurations
\begin{equation}
Z \; = \; \prod_x \sum_{p(x) \in \mathds{Z}_3 }  e^{-S[p]} \;  
= \; \sum_{\{p\}} e^{-S[p]} \;.
\label{sum_z3} 
\end{equation}
 
\section{Solving the complex phase problem}
\vspace{-1mm}
\noindent Both effective theories have complex action when $\mu \neq 0$ and thus are not
directly suitable for a Monte Carlo simulation.  Applying high temperature expansion
techniques, the partition function can be rewritten in terms of new degrees of freedom, so
called flux variables. In the flux representation the new Boltzmann factors are always real and
non-negative and a Monte Carlo simulation is possible. In this contribution we outline only the
general strategy for the derivation of the flux representation in the SU(3) case
and for the details refer to 
\cite{cg} for the SU(3) model and to \cite{flux,ours} for the $\mathds{Z}_3$ case. 
The general steps to obtain the flux representation are:
\vspace{-1mm}
\begin{itemize}
\item The first step is to write the Boltzmann weight in a factorized form and to expand 
the exponentials for individual links (nearest neighbor terms) and sites (magnetic terms).
  \begin{itemize}
  \vspace{-1mm}
  \item For the nearest neighbor term this step constitutes an expansion in $\tau$  (which is 
  equivalent to high temperature expansion in statistical mechanics because there $\tau$ should
  be identified with the inverse temperature $\beta$). 
  $$
  e^{\tau L(x)L(x+\hat{\nu})^\star}\ \rightarrow\ 
  \sum_{l_{x,\nu}}\frac{\tau^{l_{x,\nu}}}{l_{x,\nu}!} 
  \big[L(x)L(x+\hat{\nu})^\star\big]^{l_{x,\nu}}\; ; \;
  e^{\tau L(x)^\star L(x+\hat{\nu})}\ \rightarrow\ 
  \sum_{\overline{l}_{x,\nu}}\frac{\tau^{\overline{l}_{x,\nu}}}{\overline{l}_{x,\nu}!}
  \big[L(x)^\star L(x+\hat{\nu})\big]^{\overline{l}_{x,\nu}}
  $$
  \item Magnetic term 
  (we use $\eta \equiv \kappa e^\mu$ and $\overline{\eta} \equiv \kappa e^{-\mu}$):
  \vspace{-1mm}
  $$
  e^{\eta L(x)}\ \rightarrow\ \sum_{s_x}\frac{\eta^{s_x}}{s_x!} L(x)^{s_x}\;\  ; \;\ 
  e^{\overline{\eta} L(x)^\star}\ \rightarrow\ 
  \sum_{\overline{s}_x}\frac{\overline{\eta}^{\overline{s}_x}}{\overline{s}_x!} L(x)^{\star\ \overline{s}_x}
  $$
  \end{itemize}
  \vspace{-2mm}
\item Reorganizing products and sums we rewrite the partition function as:
\begin{equation}
Z \; = \; \sum_{\{l,\overline{l}\}} \sum_{\{s,\overline{s}\}}
 \left( \prod_{\overline{x},\nu} 
 \frac{\tau^{l_{x,\nu}+\overline{l}_{x,\nu}}}{l_{x,\nu}!\overline{l}_{x,\nu}!}  \right) 
 \left( \prod_x \frac{\eta^{s_x}\overline{\eta}^{\overline{s}_x}}{s_x!\overline{s}_x!} \right) 
 \left( \prod_x \int D[P] L(x)^{f(x)} L(x)^{\star \, \overline{f}(x)} \right) \; ,
\end{equation}
\vspace{-1mm}
\noindent
where $f(x) \; = \; \sum_{\nu=1}^{3} [l_{x,\nu}+\overline{l}_{x-\hat{\nu},\nu}]\ +\ s_x$ and 
$\overline{f}(x) \; = \; 
\sum_{\nu=1}^{3} [l_{x-\hat{\nu},\nu}+\overline{l}_{x,\nu}]\ +\ \overline{s}_x$ 
denote the summed fluxes at the sites $x$ of the lattice.
\vspace{-2mm}
\item The last step is to integrate out the SU(3) variables $P(x)$. 
The new form of the partition sum depends only on the flux variables:
\vspace{-2mm}
  \begin{itemize}
  \item Dimers  $l_{x,\nu}, \overline{l}_{x,\nu} \in [0,+\infty[$ , living on the links $(x,\nu)$.
\vspace{-1mm}
  \item Monomers  $s_x,\overline{s}_x \in [0,+\infty[$ , living on the sites $x$.
  \end{itemize}
\vspace{-2mm}  
\item The flux variables $l_{x,\nu}, \overline{l}_{x,\nu}, s_x, \overline{s}_x$ are the new
degrees of freedom and $\sum_{\{l,\overline{l}\}} \sum_{\{s,\overline{s}\}}$ denotes the sum
over all their configurations. The flux variables are subject to a constraint which 
forces the total flux $f(x) - \overline{f}(x)$ to be a multiple of 3 at each site $x$.
All admissible flux configurations can be shown to have a positive weight
and the complex phase problem is solved.
\end{itemize}
\vspace{-2mm}
\noindent For the $\mathds{Z}_3$ effective theory the flux representation \cite{flux}
is simpler since there is only a single dimer per link and a single monomer per site, both
with values $-1, 0$ and $+1$. 

\section{Numerical analysis}
\begin{figure}[h]
\begin{center}
\includegraphics[width=7cm,clip]{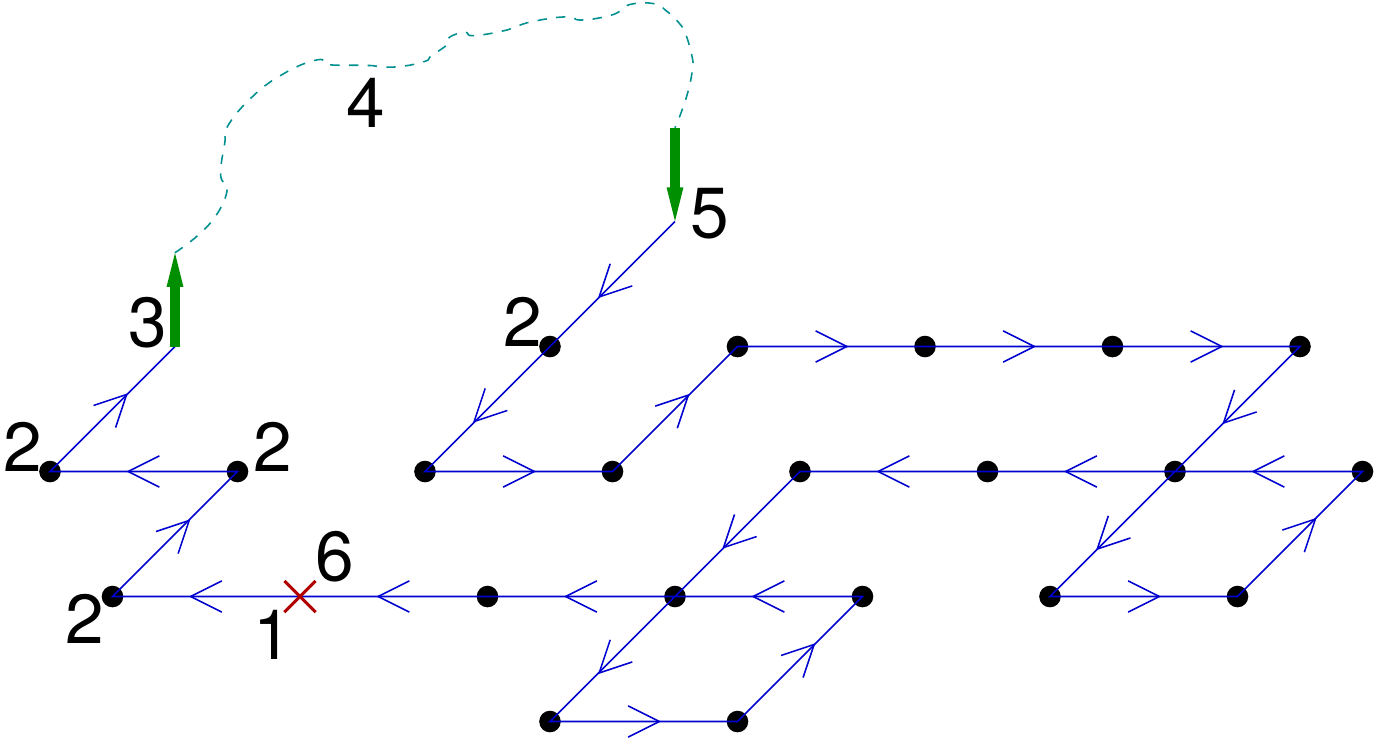}
\end{center}
\vspace{-4mm}
\caption{Schematic illustration of the worm algorithm on a 2d lattice.\vspace{-4mm}}
\label{worm1}
\end{figure}

\vspace{-1mm}
\noindent For the Monte Carlo simulation we use a generalized form of the
Prokof'ev-Svistunov worm algorithm \cite{worm} for the $\mathds{Z}_3$ effective
theory, while for the more involved SU(3) case so far only a local Metropolis
algorithm was developed. The generalization of the original Prokof'ev-Svistunov worm
algorithm \cite{worm} becomes necessary, since the constraint in the
$\mathds{Z}_3$-model enforces the conservation of flux  only modulo 3 and non-zero
monomer terms $s_x$ may give rise to additional flux at a site. Our generalization of
the original algorithm allows the worm to insert monomer flux and then to randomly hop
to another site of the lattice where it continues with the insertion of another 
monomer. It can be shown that this procedure is ergodic. The resulting
algorithm consists of four different moves which we illustrate in Fig.~\ref{worm1}:  
The worm starts at a random position (1). It may decide to insert dimer fluxes
(positions 2) but also monomers (3). The insertion of a monomer is followed by a
random hop (4) to another position, where again a monomer is  inserted (5). These
steps are continued until the worm closes (6). At each individual step the acceptance
of the proposed change is governed by a Metropolis decision. For alternative
strategies in the $\mathds{Z}_3$ model see \cite{z3old}.

\section{Results for the $\mathds{Z}_3$ effective theory}
\vspace{-1mm}
\noindent We performed several checks of the new worm algorithm:  We reproduced the
results for vanishing $\kappa$, where the theory is reduced to the 3-state
Potts model and has a first order transition at $\tau=0.183522(3)$ \cite{Karsch}.  
For small $\tau$ we calculated the partition function perturbatively up to 
${\cal O}(\tau^3)$ (dashed curves at the bottom 
of Figs.~2(a) and 3(a)).  Excellent agreement between the MC and the power series was
found.  Finally, two independent programs were written for cross checks.

\begin{figure}[ht]
\centering
\subfigure{
\includegraphics[width=0.475\textwidth,clip]{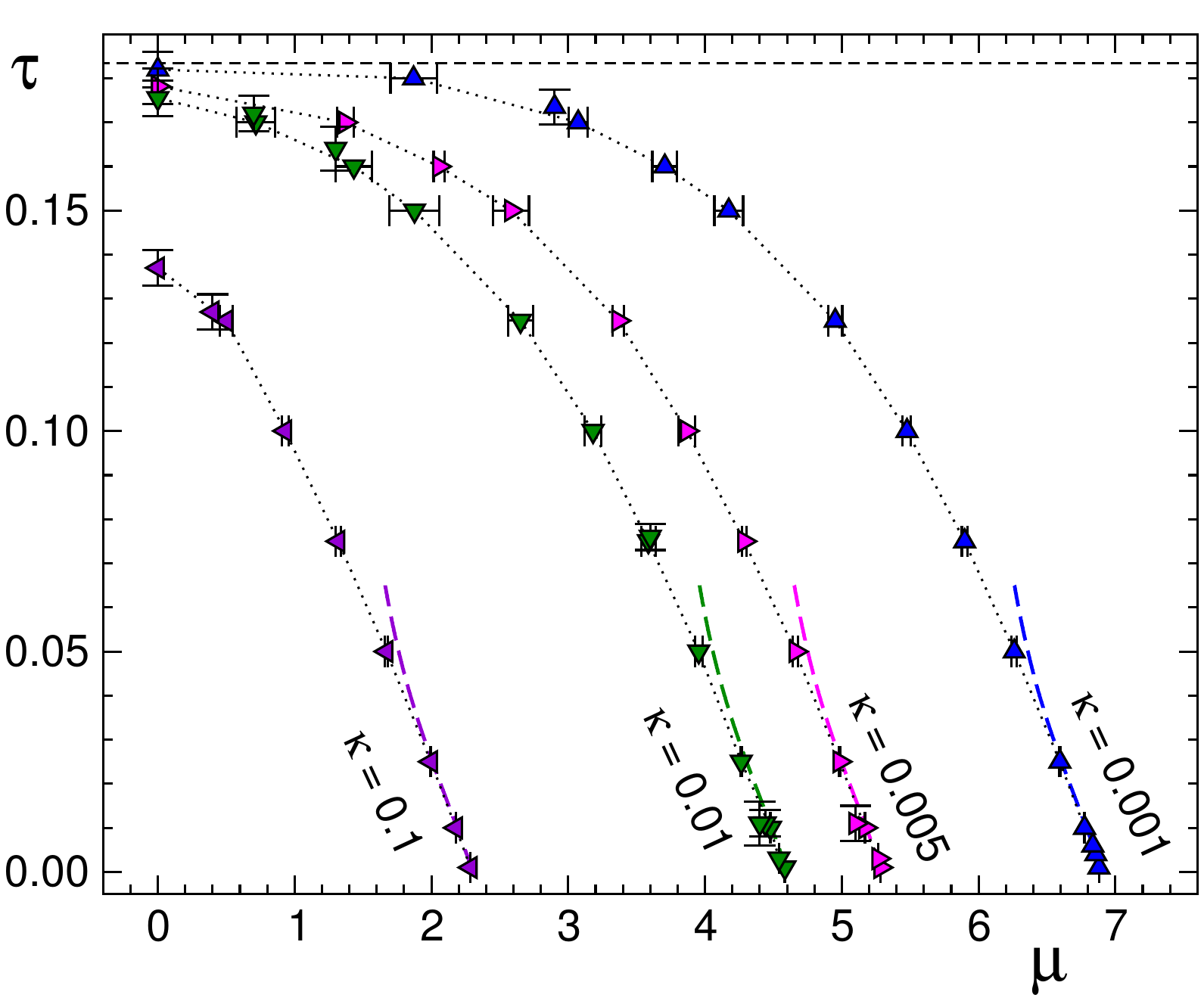}
\label{z3_xp}
}
\subfigure{
\includegraphics[width=0.475\textwidth,clip]{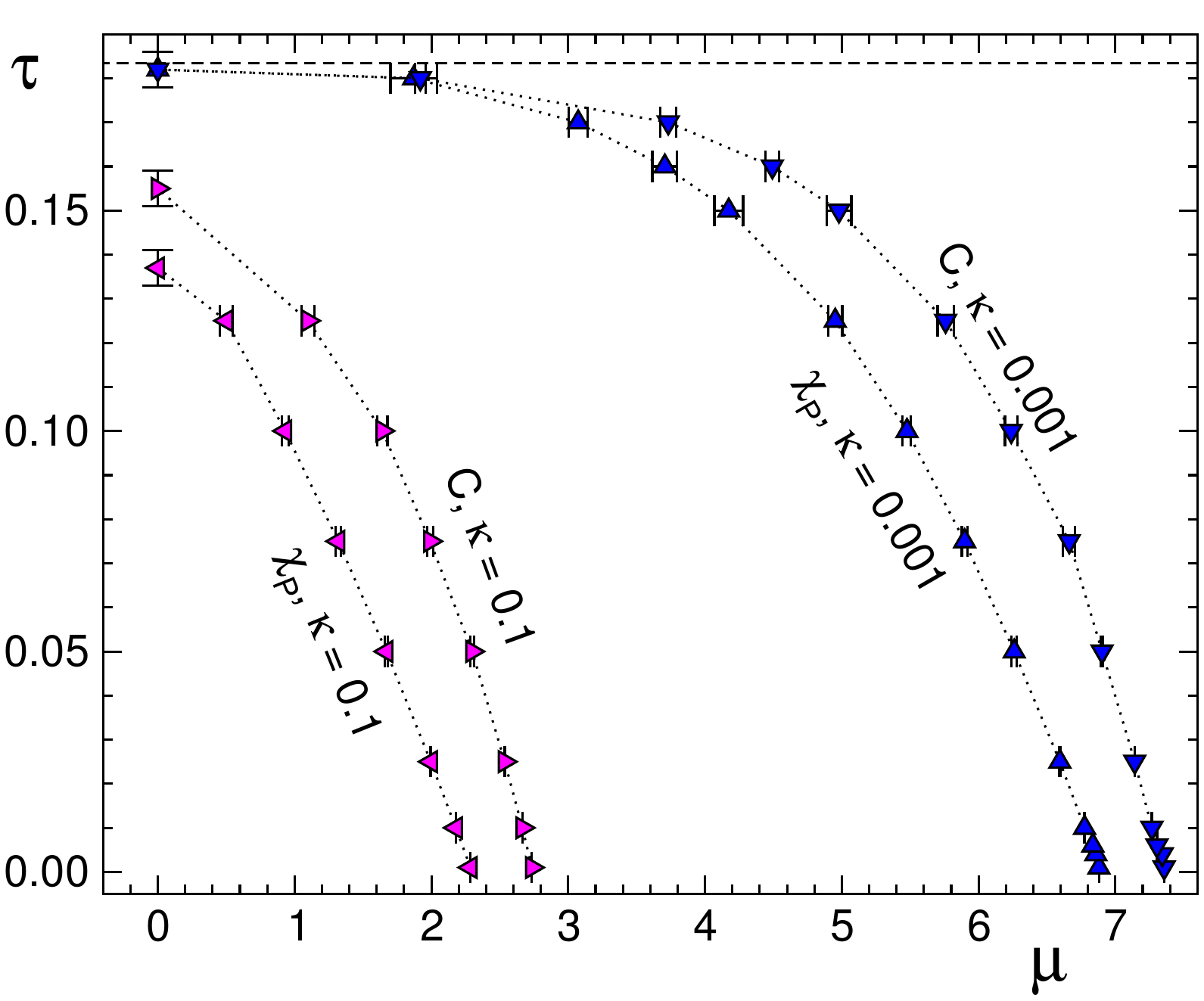}
\label{z3_comparison}
}
\vspace*{-2mm}
\caption{Phase boundaries in the $\tau$-$\mu$ plane for the $\mathds{Z}_3$ model. (a)
Left: Phase diagram obtained  from the maxima of $\chi_P$ for 4 values of $\kappa$. 
The horizontal line marks the critical $\tau$ for $\kappa = 0$. The dashed curves at the
bottom are the results from a $\tau$ expansion. (b) Right: Comparison of the phase
boundaries obtained from the maxima of the susceptibility $\chi_P$ and 
the heat capacity $C$ for two values
of $\kappa$.} \label{z3} 
\end{figure}

\vspace*{2mm}
\noindent For the analysis we focus on bulk observables and their fluctuations:
The internal energy $U$ and the magnetization $\langle p_x \rangle$,  which is
identified with the vacuum expectation value of the  Polyakov loop of QCD where a
vanishing Polyakov loop indicates confinement while a non-zero value
characterizes the deconfined phase. The corresponding fluctuations are the  heat
capacity $C$ and the Polyakov loop susceptibility $\chi_P$. All these observables can
be mapped to the flux representation where they correspond to expectation values and
fluctuations of dimers and monomers. We performed simulations on $36^3$ and $72^3$
lattices with 4 values of $\kappa$ ($0.1$, $0.01$, $0.005$ and $0.001$) and
chemical potentials up to 7.5. The phase boundaries are determined from the positions
of the maxima  of $\chi_P$ and $C$. In Fig.~\ref{z3_xp} we show the boundaries in
the $\tau-\mu$ plane as found from $\chi_P$ for all values of $\kappa$ we studied. In
Fig.~\ref{z3_comparison} we compare the boundaries from $\chi_P$ to those from $C$ for
two values of $\kappa$. It is obvious that the curves do not coincide indicating that 
the transitions into the deconfined phase are of a crossover nature. This picture was
confirmed by comparing the heights of the maxima of $\chi_P$ and $C$ for different
volumes, and the absence of a volume dependence again indicates a crossover. 

\section{Results for the SU(3) effective theory}
\vspace{-1mm}
\noindent The SU(3) effective theory has a considerably more complicated flux structure
than the $\mathds{Z}_3$ model: The number of variables is doubled  and each variable
assumes values in $[0,\infty[$. Currently we simulate the system  with a local
Metropolis update where flux around plaquettes, three units of flux on the same link, or
one unit of flux on a link with monomers on the ends are offered to change a
configuration. Again we performed several checks of our program:  We reproduced the
results for vanishing $\kappa$, where the theory can be updated in the spin
representation (\ref{action_su3}). For small $\tau$ we calculated the partition function
perturbatively taking into account terms up to $\tau^2$ (dashed curves in Fig.~3(a)).   
Finally, independent programs were written for cross checks.

\begin{figure}
\centering
\subfigure{
\includegraphics[width=0.475\textwidth,clip]{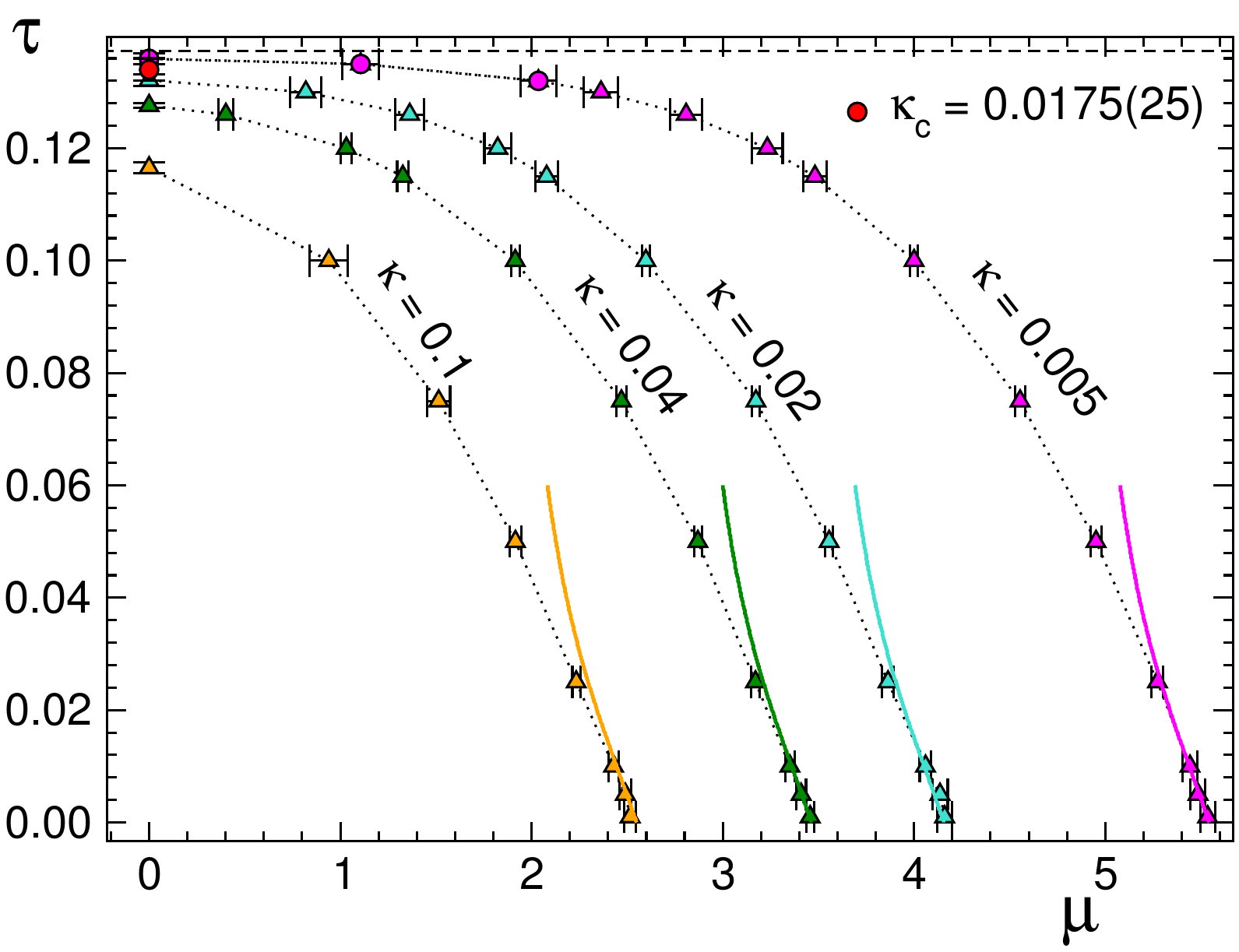}
\label{su3_xp}
}
\subfigure{                       
\includegraphics[width=0.475\textwidth,clip]{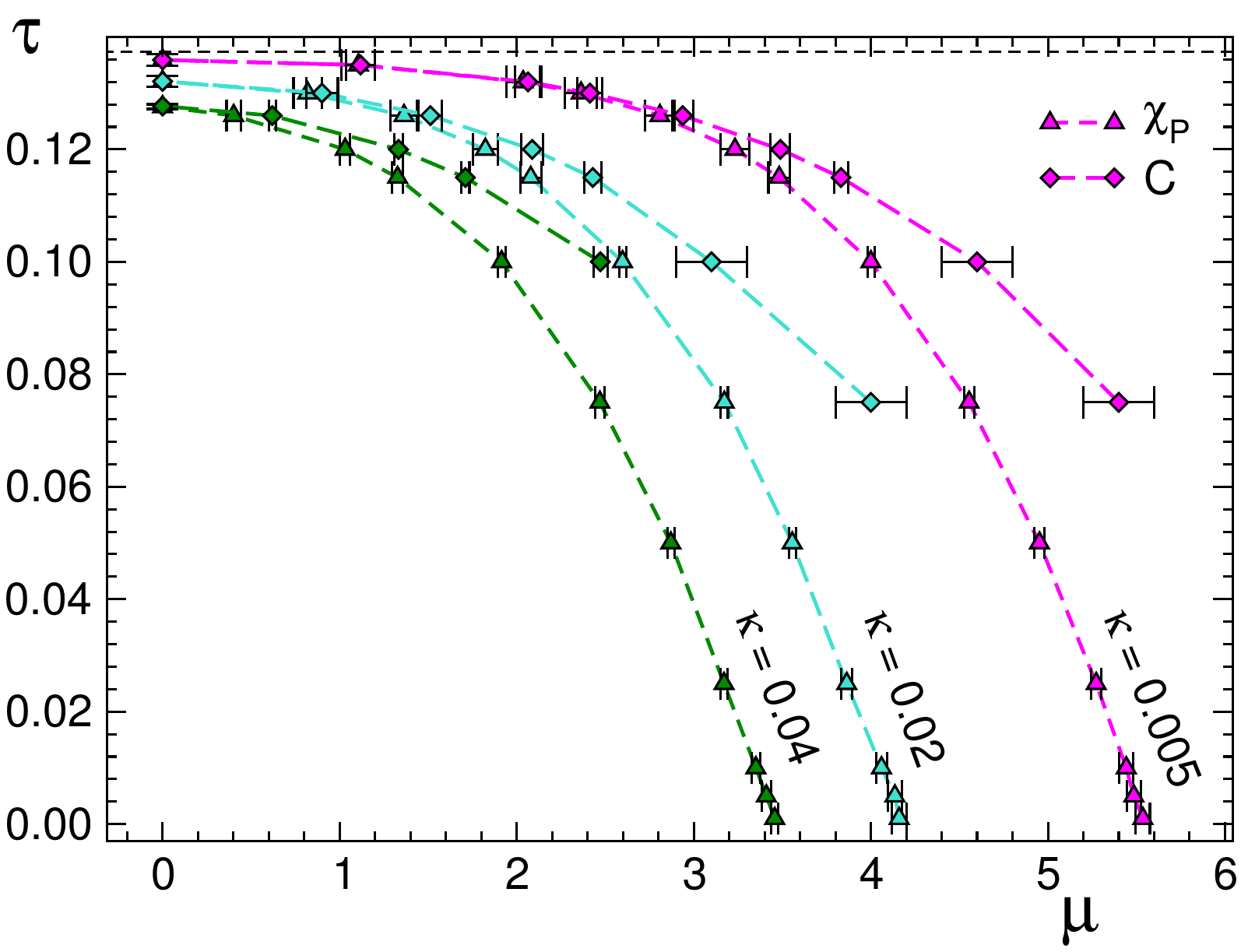}
\label{su3_comparison}
}
\vspace*{-2mm}
\caption{Phase boundaries of the SU(3) model in the $\tau$-$\mu$ plane. (a) Left:
Phase diagram obtained from the maxima of $\chi_P$ for 4 values of $\kappa$.  The
horizontal line marks the critical $\tau$ for $\kappa = 0$, and the curves at the
bottom are the results from a $\tau$ expansion. The red point is the critical end
point for $\kappa = 0$. (b) Right: Comparison of the phase boundaries obtained from
the maxima of $\chi_P$ and $C$ for three values of $\kappa$.}
\end{figure}

\vspace*{2mm} \noindent We performed simulations on $10^3$, $12^3$, $16^3$ and $20^3$ lattices
with periodic boundary conditions at finite chemical potential and $\kappa\ =$ $0.1$,
$0.04$, $0.02$ and $0.005$.  Again we identify the phase boundaries from the maxima of
$\chi_P$ and $C$. Fig.~\ref{su3_xp} shows the position of the maxima of $\chi_P$ in the
$\tau-\mu$ plane. We find that there is a first order phase transition for small
$\mu$ and $\kappa < \kappa_c$ (circles), while the rest is a crossover (triangles). 
Fig.~\ref{su3_comparison} shows the positions of the maxima of $\chi_P$ and $C$,
demonstrating that the crossover region becomes wider with increasing $\mu$. 
To determine the nature of the transitions we use two methods: First we study 
the histograms of $U$ and $P$ to check if there
is a double peak behavior characteristic of a first order transition, and, secondly, 
we analyze the volume scaling of the $C$ and $\chi_P$.  We find 
that for $\kappa \geq 0.04$ the transition is a smooth crossover at any value of
$\mu$.  For $\kappa < 0.02$ and vanishing $\mu$ the transition is of first order.
The first order behavior persists until it ends in a critical end point. To
determine the exact position of the end point and its $\mu$ dependence, we are
currently evaluating  Binder cumulants. So far we have a first estimate for the critical point for
$\mu = 0$ at  $(\tau_c,\kappa_c) = (0.130(2),0.0175(25))$.

\section{Conclusions and outlook}
\vspace{-1mm}
\noindent We have studied two effective theories of QCD with finite quark density at
non zero temperature.  Mapping the models to a flux representation enables us not
only to have a model free of the complex phase problem but also opens the possibility 
to use generalized worm algorithms for the update.   
For small values of $\kappa$ (physical case) the transition is of a smooth crossover 
type for both models
and we conclude that center symmetry alone does not provide a mechanism for first
order behavior in the QCD phase diagram. From a more technical point of view  our results constitute a
controllable reference case that can be used to test other appproaches to 
finite density lattice QCD.

\section*{Acknowledgments} 
\vspace{-1mm}
\noindent
We thank Gerd Aarts, Shailesh Chandrasekharan and Christian Lang 
for valuable discussions and remarks. This work was supported 
by the Austrian Science Fund, FWF, DK {\it Hadrons in Vacuum, Nuclei, and Stars} 
(FWF DK W1203-N16)
and by the Research Executive Agency (REA) of the European Union 
under Grant Agreement number PITN-GA-2009-238353 (ITN STRONGnet).

\end{document}